\documentclass[aps,prd,superscriptaddress,nofootinbib,amsmath,amsfonts,preprintnumbers,groupedaddress,showpacs,10pt,english]{revtex4}
\usepackage{amsmath}
\usepackage{amssymb}
\usepackage{babel}
\usepackage{wrapfig}
\usepackage{cancel}

\usepackage{relsize,exscale}
\makeatletter

\usepackage{array,multirow,graphicx}
\usepackage{dcolumn}
\usepackage{newlfont}
\usepackage{bm}
\usepackage[colorlinks,citecolor=blue,urlcolor=blue,linkcolor=blue]{hyperref}
\usepackage[figtopcap]{subfigure}
\usepackage{color}
\usepackage{verbatim}

\def\be{\begin{align}}
\def\ee{\end{align}}
\def\bea{\begin{eqnarray}}
\def\eea{\end{eqnarray}}
\def\bal{\begin{align}}
\def\eal{\end{align}}

\usepackage{scalerel}
\usepackage{tikz}
\usetikzlibrary{svg.path}
\definecolor{orcidlogocol}{HTML}{A6CE39}
\tikzset{
 orcidlogo/.pic={
 \fill[orcidlogocol] svg{M256,128c0,70.7-57.3,128-128,128C57.3,256,0,198.7,0,128C0,57.3,57.3,0,128,0C198.7,0,256,57.3,256,128z};
 \fill[white] svg{M86.3,186.2H70.9V79.1h15.4v48.4V186.2z}
 svg{M108.9,79.1h41.6c39.6,0,57,28.3,57,53.6c0,27.5-21.5,53.6-56.8,53.6h-41.8V79.1z M124.3,172.4h24.5c34.9,0,42.9-26.5,42.9-39.7c0-21.5-13.7-39.7-43.7-39.7h-23.7V172.4z}
 svg{M88.7,56.8c0,5.5-4.5,10.1-10.1,10.1c-5.6,0-10.1-4.6-10.1-10.1c0-5.6,4.5-10.1,10.1-10.1C84.2,46.7,88.7,51.3,88.7,56.8z};}}
\newcommand\orcid[1]{\href{https://orcid.org/#1}{\mbox{\scalerel*{
\begin{tikzpicture}[yscale=-1,transform shape]
\pic{orcidlogo};
\end{tikzpicture}
}{|}}}}
\graphicspath{{./}{Figs/}}
\begin{document}
\date{\today}
\title{Maxwell-$f(Q)$ theory}

 \author{G.G.L. Nashed}
\email{nashed@bue.edu.eg}
\affiliation {Centre for Theoretical Physics, The British University in Egypt, P.O. Box
43, El Sherouk City, Cairo 11837, Egypt}
\affiliation {Center for Space Research, North-West University, Potchefstroom 2520, South Africa}

%\author{Emmanuel N. Saridakis}
%\email{Emmanuel\_Saridakis@baylor.edu}
%\affiliation{Department of Physics, National Technical University of Athens, Zografou
%Campus GR 157 73, Athens, Greece}
%\affiliation{CASPER, Physics Department, Baylor University, Waco, TX 76798-7310, USA}

\begin{abstract}
Exploring the four-dimensional AdS black hole is crucial within the framework of the AdS/CFT correspondence. In this research, considering the charged scenario, we investigate the four-dimensional stationary and rotating AdS solutions in the framework of the $f(Q)$ gravitational theory. Our emphasis is on the power-law ansatz, which is consistent with observations and is deemed the most viable. Because this solution does not have an uncharged version or relate to general relativity, it falls into a new category, which derives its features from changes in non-metricity and incorporates the Maxwell domain. We analyze the singularities of such a solution, computing all the quantities of different curvature and non-metricity invariants.  Our results indicate the presence of a central singularity, albeit with a softer nature compared to standard non-metricity or Einstein general relativity, attributed to the influence of the effect of $f(Q)$.  We examine several physical characteristics of black holes from a thermodynamics perspective and demonstrate the existence of an outer event horizon in addition to the inner Cauchy horizons. However, under the conditions of a sufficiently large electric charge, a naked singularity emerges. Finally, we derive a class of rotating black hole in four-dimensional $f(Q)$ gravity that are asymptotically anti-de Sitter charged.
%\pacs{04.50.Kd,  98.80.âk, 97.60.Lf}
\end{abstract}

\maketitle

%%%%%%%%%%%%%%%%%%%%%%%%%%%%%%%%%%% Section 1 %%%%%%%%%%%%%%%%%%%%%%%%%%%%%%%%%%%%%%%%
\section{Introduction}\label{S1}
%%%%%%%%%%%%%%%%%%%%%%%%%%%%%%%%%%%%%%%%%%%%%%%%%%%%%%%%%%%%%%%%%%%%%%%%%%%%%%%%%%%%%%

Over the past few years, there has been considerable debate about gravitational theories that incorporate non-metricity~\cite{Nester:1998mp,BeltranJimenez:2018vdo,BeltranJimenez:2019esp,Runkla:2018xrv,BeltranJimenez:2019tme}.
These gravitational theories use a quantity called the non-metricity scalar $Q$ and follow the Palatini formalism. This approach treats the connection, along with the metric, as independent variables. When the Riemann tensor and torsion tensor for the connection are set to zero, it is found that the connection can be described using four scalar fields~\cite{Blixt:2023kyr,BeltranJimenez:2022azb,Adak:2018vzk,Tomonari:2023wcs}.
Furthermore, by choosing certain scalar fields, one can set the connection to zero, a condition known as the coincident gauge \cite{DAmbrosio:2021zpm}. In this scenario, the metric becomes the sole dynamic variable in the theory. It has been shown that in the coincident gauge, a theory with a linear action, denoted as $Q$, is equivalent to general relativity (GR). This equivalence is termed as Symmetric Teleparallel Equivalent to GR (STEGR)~\cite{BeltranJimenez:2017tkd,Quiros:2021eju}.

Teleparallelism is a widely recognized concept in gravitational theory that operates independently of curvature. In this framework, the torsion scalar $T$ is considered as the fundamental quantity~\cite{Maluf:2013gaa,Aldrovandi:2013wha,Cai:2015emx}.
Choosing the Weitzenb\"ock connection and using the coincident gauge in STEGR results in the elimination of the spin connection, leaving only the tetrad as the dynamic variable. The theory with an action linear to $T$ is equivalent to GR and is called teleparallel equivalent to GR (TEGR)~\cite{Maluf:2013gaa,Krssak:2015oua}. In non-metricity gravity, unlike teleparallelism, we use ``symmetric" teleparallelism to meet the torsionless requirement, which requires a symmetric affine connection. Similarly to how we extend GR to $f(R)$ gravity \cite{Nashed:2018piz}, which involves a function of the curvature scalar, TEGR \cite{Nashed:2018qag,Nashed:2020kdb}, and STEGR have also been extended to $f(T)$ and $f(Q)$ gravity, respectively. These extensions incorporate arbitrary functions of $T$ and $Q$ in their actions~\cite{Bahamonde:2015zma,Cai:2015emx,Krssak:2015oua,Bahamonde:2021gfp,Hu:2023gui,Heisenberg:2018vsk,BeltranJimenez:2019tme,Harko:2018gxr,Jarv:2018bgs,Runkla:2018xrv,Capozziello:2022zzh,Capozziello:2023vne}.
These theories have been extensively studied in the context of modified gravity. By introducing new degrees of freedom (DOF) through torsion or non-metricity, a range of fascinating phenomena has been revealed, especially in cosmological models~\cite{Hu:2023ndc,Qiu:2018nle,Najera:2021afa,Li:2021mdp,Khyllep:2021pcu,Casalino:2020kdr,Sharma:2021fou,Sharma:2021ivo,Sharma:2021ayk,Mandal:2020buf,Mandal:2020lyq,Mandal:2021bpd,Arora:2022mlo,Gadbail:2022jco,
Paliathanasis:2023kqs,Paliathanasis:2023nkb,Dimakis:2022rkd,Paliathanasis:2023gfq}, studies have explored solutions related to black holes~\cite{Wang:2021zaz,Lin:2021uqa,DAmbrosio:2021zpm,Bahamonde:2022esv,Calza:2022mwt}, as well as gravitational waves~\cite{Hohmann:2018jso,Hohmann:2018wxu,Soudi:2018dhv,Abedi:2017jqx}. { In \cite{DAmbrosio:2021zpm} authors derived many perturbative spherically symmetric solutions using different forms of $f(Q)$.}

Recently, there has been growing interest in the number of DOF in $f(Q)$ gravity~\cite{Hu:2022anq,DAmbrosio:2023asf,Tomonari:2023wcs,Heisenberg:2023lru,Paliathanasis:2023pqp,Dimakis:2021gby}. In a prior investigation by Hu et al.~\cite{Hu:2022anq}, they showed that coincident $f(Q)$ gravity has eight DOFs, with the scalar mode unable to propagate.  However, it's important to mention that the conclusions from Ref.~\cite{Hu:2022anq} have sparked controversy and are currently being actively discussed (for more details, c.f.~\cite{DAmbrosio:2023asf,Tomonari:2023wcs}). { However, the current literature concerning Hamiltonian analysis heavily relies on a specific gauge called the coincident gauge \cite{DAmbrosio:2023asf}.} It is essential to confirm that there are no ghost modes with negative kinetic energy among the physical DOFs without resorting to gauge fixing. This study primarily concentrates on a charged black hole solution within the cubic form of $f(Q)$ theory. Additionally, we examine the construction of this solution by studying its invariants as well as thermodynamic quantities.

The structure of this paper is outlined as follows:
In Section~\ref{S1}, we introduce the basic geometrical framework of charged $f(Q)$ gravitational theory.  In Section~\ref{AdsSection}, we employ cylindrical coordinates (t, r, $\xi_1$, and $\xi_2$) to apply the charged field equation of $f(Q)$ to a four-dimensional line element with two unknown functions.  In Section~\ref{AdsSection}, we present both charged and uncharged black holes within the cubic form of $f(Q)$. Furthermore, we discuss the interaction between the physics of the charged black hole and its singularity in the same section. In Section~\ref{S5}, we propose a new rotating black hole using the cubic form of the $f(Q)$ gravitational theory. Section~\ref{S6} is dedicated to calculate the entropy, Hawking temperature, and heat capacity to analyze the thermodynamic behavior of the charged black hole. Finally, in Section~\ref{S7}, we conclude this study and discuss the findings of our investigation.

%%%%%%%%%%%%%%%%%%%%%%%%%%%%%%%%%%% Section 1 %%%%%%%%%%%%%%%%%%%%%%%%%%%%%%%%%%%%%%%%
\section{Maxwell-$f(Q)$ theory}
\label{S1}

Weyl geometry represents a substantial advancement beyond Riemannian geometry, forming the mathematical underpinning of GR. In Weyl geometry, when a vector undergoes parallel transport along a closed path, it not only alters its direction but also its magnitude. Consequently, in the framework of Weyl's the metric tensor's covariant derivative is not zero, and this feature can be mathematically represented by a recently introduced geometric quantity known as non-metricity, which is represented by the letter $Q$.  Hence, the covariant derivative of the metric tensor $g_{\mu \nu }$ with respect to the general affine connection, $\overline{\Gamma }_{~\mu \gamma }^{\sigma }$, yields the definition of the non-metricity tensor $Q_{\gamma \mu \nu }$, and it can be formulated as \cite{BeltranJimenez:2017tkd,BeltranJimenez:2019tme, BeltranJimenez:2018vdo},%
\begin{equation}
Q_{\gamma \mu \nu }=-\nabla _{\gamma }g_{\mu \nu }=-\frac{\partial g_{\mu
\nu }}{\partial x^{\gamma }}+g_{\nu \sigma }\overline{\Gamma }_{~\mu \gamma
}^{\sigma }+g_{\sigma \mu }\overline{\Gamma }_{~\nu \gamma }^{\sigma }.
\label{2f}
\end{equation}

Under these circumstances, the Weyl connection that characterizes the general affine connection can be divided into two separate parts, which are as follows:
\begin{equation}\label{1}
\overline{\Gamma }_{\ \mu \nu }^{\gamma }=\left\{ {}^{\, \alpha}_{\mu \nu} \right\}+L_{\ \mu \nu }^{\gamma }.
\end{equation}%
In Eq. (\ref{1}),  the Levi-Civita connection for the metric $g_{\mu \nu }$ is represented by the first term which and defined as:
\begin{equation}
\left\{ {}^{\, \alpha}_{\mu \nu} \right\}\equiv\frac{1}{2}g^{\gamma \sigma }\left( \frac{%
\partial g_{\sigma \nu }}{\partial x^{\mu }}+\frac{\partial g_{\sigma \mu }}{%
\partial x^{\nu }}-\frac{\partial g_{\mu \nu }}{\partial x^{\sigma }}\right)
.
\end{equation}

Moreover,  the second term in Eq. (\ref{1}) expresses the tensor of disformation resulting from the non-metricity  that can be expressed in the following manner:
\begin{equation}
L_{~\mu \nu }^{\gamma }\equiv\frac{1}{2}g^{\gamma \sigma }\left( Q_{\nu \mu
\sigma }+Q_{\mu \nu \sigma }-Q_{\gamma \mu \nu }\right) =L_{~\nu \mu
}^{\gamma }.
\end{equation}%
Moreover, by contracting the non-metricity tensor we derive the non-metricity scalar from the disformation tensor  as:
\begin{equation}
Q\equiv -g^{\mu \nu }\left( L_{\ \ \beta \mu }^{\alpha }L_{\ \ \nu \alpha
}^{\beta }-L_{\ \ \beta \alpha }^{\alpha }L_{\ \ \mu \nu }^{\beta }\right) .
\end{equation}

Expansion of Weyl geometry to include space-time torsion results in the Weyl-Cartan geometries having torsion. In the geometry of Weyl-Cartan, the general affine connection could be divided into two separate components:
\begin{equation}\label{2}
\overline{\Gamma }_{\ \mu \nu }^{\gamma }=\left\{ {}^{\, \alpha}_{\mu \nu} \right\}+K_{\ \mu \nu }^{\gamma }.
\end{equation}%
In Eq. (\ref{2}), the part on the right-hand side shows  contortion  which is described by  the torsion tensor
 $T_{\ \mu \nu }^{\gamma }$ It is defined as $2\overline{\Gamma }_{~[\mu \nu ]}^{\gamma }$ and can have the following form:
\begin{equation}
K_{\ \mu \nu }^{\gamma }\equiv \frac{1}{2}g^{\gamma \sigma }\left( T_{\mu
\sigma \nu }+T_{\nu \sigma \mu }+T_{\sigma \mu \nu }\right) .
\end{equation}

Moreover, the connection between the curvature tensors $R_{\sigma \mu \nu }^{\rho }$ and $\mathring{R}{\sigma \mu \nu }^{\rho }$ associated with the connections $\overline{\Gamma }{\ \mu \nu }^{\gamma }$ and $\Gamma _{\ \mu \nu }^{\gamma }$ is:%
\begin{equation}
R_{\sigma \mu \nu }^{\rho }=\mathring{R}_{\sigma \mu \nu }^{\rho }+\mathring{%
\nabla}_{\mu }L_{\nu \sigma }^{\rho }-\mathring{\nabla}_{\nu }L_{\mu \sigma
}^{\rho }+L_{\mu \lambda }^{\rho }L_{\nu \sigma }^{\lambda }-L_{\nu \lambda
}^{\rho }L_{\mu \sigma }^{\lambda },
\end{equation}%
\begin{equation}
R_{\sigma \nu }=\mathring{R}_{\sigma \nu }+\frac{1}{2}\mathring{\nabla}_{\nu
}Q_{\sigma }+\mathring{\nabla}_{\rho }L_{\nu \sigma }^{\rho }-\frac{1}{2}%
Q_{\lambda }L_{\nu \sigma }^{\lambda }-L_{\sigma \lambda }^{\rho }L_{\rho
\sigma }^{\lambda },  \label{2g}
\end{equation}%
and the relationship regarding Ricci scalar  has the from:%
\begin{equation}
R=\mathring{R}+\mathring{\nabla}_{\lambda }Q^{\lambda }-\mathring{\nabla}%
_{\lambda }\tilde{Q}^{\lambda }-\frac{1}{4}Q_{\lambda }Q^{\lambda }+\frac{1}{%
2}Q_{\lambda }\tilde{Q}^{\lambda }-L_{\rho \nu \lambda }L^{\lambda \rho \nu
}.  \label{2h}
\end{equation}%
withe $\mathring{\nabla}$ represents the operator of the covariant derivative associated with the Levi-Civita relationship $\Gamma _{\ \mu \nu }^{\gamma }$.  The general affine connection in STEGR is constrained by the lack of curvature and torsion requirements. For a Riemann tensor $R_{\sigma \mu \nu }^{\rho }\left( \overline{\Gamma }\right)$ to be curvature-free  it must has a vanishing value.  The parallel transport performed by the covariant derivative $\nabla$ and its affine connection $\overline{\Gamma }_{\ \mu \nu }^{\gamma }$ becomes path-independent when the Riemann tensor vanishes.  This theory, STEGR, implies that $T_{\ \mu \nu }^{\gamma }=0$ and that the connection must be torsionless in addition to requiring zero curvature. In  this theory gravitational effects are entirely attributed to the  non-metricity. when the torsion tensor vanishes this yields  to symmetric lower indices for the general affine connection.

{ Now if we start from a teleparallel condition, which corresponds to a variety with a plane geometry characterizing a pure inertial connection, it is possible to perform a gauge transformation of the linear group ${ GL }(4, {{R}})$ parameterized by $\Lambda_{\phantom{\alpha}\mu}^{\alpha}$ \cite{BeltranJimenez:2019esp,BeltranJimenez:2019tme},
\begin{equation}
\Gamma_{\phantom{\alpha}\mu\nu}^{\alpha}=\left(\Lambda^{-1}\right)_{\phantom{\alpha}\beta}^{\alpha}\partial_{[\mu}\Lambda_{\phantom{\alpha}\nu]}^{\beta}.
\end{equation}
So we can write that the most general possible connection, through the general element of ${\textrm GL}(4, {{R}})$, which is parameterized by the transformation of
$\Lambda_{\phantom{\alpha}\mu}^{\alpha}=\partial_{\mu}\xi^{\alpha}$, where $\xi^{\alpha}$ is an arbitrary vector field,
\begin{equation}
\Gamma_{\phantom{\alpha}\mu\nu}^{\alpha}=\frac{\partial x^{\alpha}}{\partial\xi^{\rho}}\partial_{\mu}\partial_{\nu}\xi^{\rho}. \label{coinc}
\end{equation}
This result shows us that the connection can be removed by a coordinate transformation. The transformation that results in the connection \eqref{coinc} being removed is called gauge coincident  \cite{BeltranJimenez:2018vdo}.
\par
Consequently, from the coincident gauge we have that the non-metricity tensor defined by \eqref{2f} becomes,
\begin{equation}
 Q_{\beta\mu\nu}\equiv \partial_{\beta}g_{\mu\nu}. \label{tns_nmetric2}
\end{equation}
In this study, we use the coincident gauge to compute our solution.}

Finally, the gravity action of $f(Q)$ can be constructed as follows using the non-metricity scalar \cite{BeltranJimenez:2017tkd}:
\begin{equation}\label{action1}
    S=-\frac{1}{2\kappa^2}\int_\mathcal{M}   f({Q}) \sqrt{-g} d^4x +\int \sqrt{-g} {\cal L}_{ em}~d^{4}x \,.
\end{equation}
Here, the space-time manifold, the covariant metric tensor, and the determinant are denoted by $g$, $g_{\mu\nu}$, and $\mathcal M$, respectively. The generic functional form of the non-metricity scalar ${Q}$ is represented by the function $f({Q})$. Here $\kappa = 8 \pi$ defines $\kappa$ in relativistic units, i.e., $c=G=1$, where the gravitational constant and the speed of light are identical.   The Maxwell field Lagrangian is represented by  ${\cal L}_{
em}=-\frac{1}{2}{ F}\wedge ^{\star}{F}$  in Eq.  (\ref{action1}), where $F = dA$ and  $A=A_{\mu}dx^\mu$ is the 1-form of the electromagnetic potential \cite{Awad:2017tyz}.
Just like  $f(R)$ gravity, $f(Q)$ gravity also results in deviations from Einstein GR.  As an illustration, by setting the   $f(Q)=Q$, we recover the STEGR. From the non-metricity tensor, we can extract only two independent traces because of the symmetry of the metric tensor $g_{\mu \nu}$,
$Q_{\gamma \mu \nu }$,%
\begin{equation}
Q_{\gamma }\equiv Q_{\gamma \ \ \ \mu }^{\ \ \mu },\ \ \ \ \tilde{Q}_{\gamma
}\equiv Q_{\ \ \gamma \mu }^{\mu }.  \label{2l}
\end{equation}

Furthermore, it would be helpful to present the conjugate defining non-metricity as:

\begin{equation}
\hspace{-0.5cm}P_{\ \ \mu \nu }^{\gamma }\equiv \frac{1}{4}\bigg[-Q_{\ \ \mu
\nu }^{\gamma }+2Q_{\left( \mu \ \ \ \nu \right) }^{\ \ \ \gamma }+Q^{\gamma
}g_{\mu \nu }-\widetilde{Q}^{\gamma }g_{\mu \nu }-\delta _{\ \ (\mu
}^{\gamma }Q_{\nu )}\bigg]=-\frac{1}{2}L_{\ \ \mu \nu }^{\gamma }+\frac{1}{4}%
\left( Q^{\gamma }-\widetilde{Q}^{\gamma }\right) g_{\mu \nu }-\frac{1}{4}%
\delta _{\ \ (\mu }^{\gamma }Q_{\nu )}.
\end{equation}

The non-metricity scalar is calculated as follows:
\begin{equation}\label{Qs}
Q=-Q_{\gamma \mu \nu }P^{\gamma \mu \nu }=-\frac{1}{4}\big(-Q^{\gamma \nu
\rho }Q_{\gamma \nu \rho }+2Q^{\gamma \nu \rho }Q_{\rho \gamma \nu
}-2Q^{\rho }\tilde{Q}_{\rho }+Q^{\rho }Q_{\rho }\big).
\end{equation}

{When deriving the field equations of the theory, one performs separate variations with respect to the metric and  matter fields  of Eq.~(\ref{action1}) that yield \cite{Heisenberg:2023lru}:}
\begin{align}\label{1st EOM}
&\zeta_{\mu \nu}=\frac{2}{\sqrt{-g}} \nabla_\alpha \left( \sqrt{-g} f_{Q} P^\alpha_{\;\; \mu \nu }\right) + \frac{1}{2} g_{\mu \nu} f + f_{Q} \left( P_{\mu \alpha \beta } {Q}^{\;\;\alpha \beta}_\nu
- 2P_{\alpha \beta \mu} {Q}^{\;\;\alpha \beta}_\nu\right)+\kappa^2\frac{1}{2}\kappa{{{\cal
T}^{{}^{{}^{^{}{\!\!\!\!\scriptstyle{em}}}}}}}_{\mu \nu}\nonumber\\
&\partial_\nu \left( \sqrt{-g} F^{\mu \nu} \right)=0\;,
\end{align}
{ The variation of Eq.~(\ref{action1}) with respect to the connection yields:}
%An extra constraint on the connection can be achieved with the aid of Eq. (\ref{action1}), that gives
\begin{equation}\label{2nd EOM}
{ \nabla^\mu \nabla^\nu \left(\sqrt{-g} f_{Q} P^\alpha_{\;\; \mu \nu }\right)=0}\,.
\end{equation}
In this study, ${\mathcal{T}_\mu^\nu{}^{^\text{em}}}$ denotes the tensor characterizing the electromagnetic field's energy-momentum, which is figured as:
\[
{{{\cal
T}^{{}^{{}^{^{}{\!\!\!\!\scriptstyle{em}}}}}}}^\nu_\mu=F_{\mu \alpha}F^{\nu \alpha}-\frac{1}{4} \delta_\mu{}^\nu F_{\alpha \beta}F^{\alpha \beta}.\]
In Eq. (ref{1st EOM}),  $f$ is defined as $f({Q})$, and $f_{Q}$ is its first derivative with respect to ${Q}$. It is worth noting that the matter  Lagrangian density is varied independently regarding the connection, resulting in the absence of hyper-momentum. Furthermore, as is widely recognized, the outcomes of GR (in the framework of scalar-tensor extended GR, STEGR) are obtained by setting $f({Q})={Q}$. Consequently, the Lagrangian density takes the form $\mathcal L=-\frac{{Q}}{2\kappa^2}+\mathcal L_m$, where  $f_{Q}=\frac{df}{dQ}$.

%Similarly, without hyper-momentum \cite{BeltranJimenez:2018vdo}, the gravitational action (\ref{action1}) is varied with respect to the connection field to derive the connection field equations as:
%\begin{equation}
%\nabla _{\mu }\nabla _{\nu }(\sqrt{-g}f_{Q}P^{\mu \nu }{}_{\gamma })=0.
%\label{2q}
%\end{equation}

\section{Solution of anti-de-Sitter black hole}
\label{AdsSection}
Next, we will derive a solution of  AdS charged black hole  within  $f(Q)$ theory, with a specific emphasis on four-dimensional spacetime. We utilize cylindrical coordinates ($t$, $r$, $\xi_1$, $\xi_2$), { with the ranges \textrm{$0\leq r< \infty$, $-\infty < t < \infty$, and  $0\leq \xi_{i}< 2\pi$, $i=1,2$ }}. Within this context, let's use the following metric:
\begin{align}
\label{m2}
ds^2=
S(r)dt^2-\frac{1}{S_1(r)}dr^2-r^2\left(d\xi^2_1+d\xi^2_2\right)\,.
\end{align}
The radial coordinate $r$ is the only variable that fixes the functions $S(r)$ and $S_1(r)$.
Notably, the metric (\ref{m2}) includes flat sections rather than spherical or hyperbolic ones, indicating that it is not entirely general.
%\footnote{{ It is important to stress the fact that using the gauge coincident with the spacetime (\ref{m2}) results in the vanishing of the off-diagonal component identically. However, if the form of the metric (\ref{m2}) were changed to other coordinates, the off-diagonal component of the field equations (\ref{1st EOM}) might survive. Moreover, the procedure studied in \cite{DAmbrosio:2021zpm} also needs to be checked for the metric (\ref{m2}). All these issues are outside the scope of the present study and will be studied elsewhere.}}
 {It is shown in \cite{DAmbrosio:2021zpm}, that if we start from a general metric-affine geometry, one can construct the most general static and
spherically symmetric forms of the metric and the affine connection. Then the uses of these symmetry-reduced
geometric objects to prove that the field equations of $f(Q)$ gravity admit GR solutions and other solutions beyond GR. This is contrary to what has been known  in the literature. In this study we focus on the metric (\ref{m2})  to obtain a solution that is different from GR  where $Q$ has a non-constant value and $f(Q)$ is not linear.}

For this reason, we focus on the metric (\ref{m2}) in order to obtain novel solutions where we can derive the non-constant value of $Q$ and the non-linear form of $f(Q)$.
When we plug this metric form into the equation for non-metricity, we get:
\begin{equation}\label{df1}
Q=-{\frac {2S_1  \left( rS'+S  \right) }{{r}^{2}S }}.
\end{equation}
In this context, $S'\equiv\frac{dS}{dr}$ and $S'_1(r)\equiv \displaystyle\frac{dS_1(r)}{dr}$, and going forward, we will use the following abbreviation  $S$, $S_1$, $S'$, and $S'_1$. Ultimately, based on the best agreement with cosmological evidence \cite{Mandal:2023cag,Calza:2022mwt}, the power-law of $f(Q)$ theory is the one we analyze in the following
\begin{equation}\label{powellaw}
 f(Q)=Q+\gamma Q^2+\gamma_1 Q^3-2\Lambda,
\end{equation}
where the dimensional parameters that defined the model in the present study are $\gamma$ and $\gamma_1$. For the sake of completeness, we have additionally inserted the cosmological constant.

\subsection{AdS black hole that is asymptotically static}\label{S2}

We start in this subsection by finding static $AdS$ black hole, particularly when the electromagnetic part is non-vanishing, so we're paying attention to that  ${{{\mathfrak
T}^{{}^{{}^{^{}{\!\!\!\!\scriptstyle{em}}}}}}}^\alpha_\beta\neq 0$. In such situation,  by substituting the metric (\ref{m2}) in Eqs.~(\ref{1st EOM}), we obtain the following non-zero components:
%\newpage
\begin{eqnarray}\label{df2}
& &
\!\!\!\!\!\!\!
\zeta^r{}_r\equiv \frac {1}{{r}^{6} S^{3}}\left\{20\,\gamma_1\, S_1{}^{3}{r}^ {3} S'^{3}-6\,{r}^{2} S_1{}^{2}S  \left(\gamma {r}^{2} -10 \,\gamma_1\,S_1  \right)  S'^{2}+rS_1 S^{2} \left( 60\,\gamma_1\, S_1{}^{2}-12\,{r}^{2}\gamma\,S_1 +{r}^{4} \right) S' - S^{2} \left(  q'^{2}S_1 {r}^{6}\right.\right.\nonumber\\
&&\left.\left.+S \left( 6\,{r}^{2}\gamma\, S_1^{2}-{r}^{4}S_1 -20\,\gamma_1\,S_1{}^{3}+{\Lambda}\,{r}^{6} \right) \right)\right\}=0 \, ,\nonumber\\
& &\!\!\!\!\!\!\!\!  \zeta^{\xi_1}{}_{\xi_1}\equiv
\zeta^{\xi_2}{}_{\xi_2}=\frac{1}{{r}^{6} S^{4}} \left\{ 2\,S_1  {r}^{2}S   \left( 36\,\gamma_1\, S_1^{2}{r}^{2} S'^{2}-8\,rS_1  S   \left( \gamma\,{r}^{2} \right) S'  -12\,\gamma_1\,S_1 + S^{2} \left( 60\,\gamma_1\, S_1^{2}-12\,{r}^{2}\gamma\,S_1  +{r}^{4} \right)  \right) S''  \right.\nonumber\\
&&\left.-60\,{r}^{4} S'^{4}\gamma_1\, S_1^{3}+4 \,S \left(  15\,r\gamma_1\,S'_1  -{40}\,\gamma_1\,S_1 +3\gamma \,{r}^{2} \right) S_1^{2}{r}^{3} S'^{3}- S^{2} \left[  \left(  12\,{r}^{3}\gamma -240\,r\gamma_1\,S_1 \right) S'_1  +204\,\gamma_1\, S_1^{2}- 12\,{r}^{2}\gamma\,S_1  \right.\right.\nonumber\\
&&\left.+{r}^{4} \right] S_1  {r}^{2} S'^{2 }+ S^{3}r \left( r \left( {r}^{4}+ 300\,\gamma_1\, S_1^{2}-36\,{r}^{2 }\gamma\,S_1   \right) S'_1 +8\,{r}^{2}\gamma\, S_1^{2}+2 \,{r}^{4}S_1  -264\,\gamma_1\, S_1^{3} \right) S'  -2 S^{3} \left[2S \left( {\Lambda}\,{r}^{6}-6\,{r}^{2}\gamma\, S_1^{2}\right. \right.\nonumber\\
&&\left.\left.\left. +40\,\gamma_1\, S_1^{3} \right) -\,rS   \left( 60\,\gamma_1 S_1^{2}-12\,{r}^{2}\gamma\,S_1  +{r}^{4} \right) S'_1-2q'^{2}S_1  {r}^{6}  \right]  \right\}
=0, \nonumber\\
& &
\!\!\!\!\!\!\!
\zeta^t{}_t\equiv \frac {1}{{r}^{6} S^{3}}\left\{8\,S S_1{}^{2 }{r}^{2} \left[ 6\,\gamma_1\,S_1 rS' +S  \left( 6\,\gamma_1\,S_1 -\gamma\,{r}^{2} \right)  \right] S'' -40\,\gamma_1\, S_1{} ^{3}{r}^{3} S'^{3}+6\,{r}^{2} S_1{}^ {2}S  \left(  \gamma\,{r}^{2}-10\,\gamma_1\,S_1+10\,r\gamma_1\,S'_1 \right)  S'^{2}\right.\nonumber\\
&&\left.-12\, S^{2}S_1 r \left[ \left( {r}^{3}\gamma -10\,r\gamma_1\,S_1 \right) S'_1 +8\,\gamma_1\, S_1^{2} \right] S' - S^{2} \left[q'^{2}{ S_1} {r}^{6} -rS \left( 60\,\gamma_1\, S_1{}^{2}-12 \,{r}^{2}\gamma\,S_1 +{r}^{4} \right) S'_1 +S  \left( {\Lambda}\,{r}^{6}-{r}^{4}{ S_1}  \right.\right.\right.\nonumber\\
&&\left.\left.\left.-10\,{r}^{2}\gamma\, S_1{}^{2}+76\,\gamma_1\, S_1{}^{3} \right)  \right]\right\}
=0\,,\nonumber\\
\end{eqnarray}
{ where $q\equiv q(r)$ is an unknown function of the radial coordinate $r$ that is constructed from the ansatz of the vector potential
as:
\begin{align}
A = q(r)dt.
\end{align}}
The $\zeta^r{}_r$-component can be rewritten as
\begin{align}\label{333}
\zeta^r{}_r\equiv Q+3\gamma Q^2+5\gamma_1 Q^3+2\Lambda=0\,.
\end{align}
Equation \eqref{333} represents a third-order algebraic equation   in $Q$ indicating that $Q = Q_0 = const.$ 
Therefore, Eq.~(\ref{df1}) for $Q= const.$ straightforwardly yields the general solution \cite{Nashed:2023tua,sym16020219}, when ${{{\mathfrak T}^{{}^{{}^{^{}{\scriptstyle{em}}}}}}}^\nu_\mu= 0$, i.e. $q(r)=0$, in the form:
\begin{eqnarray}\label{df4g}
& &  S(r)= \Lambda_{eff}r^2-\frac{M}{r},\nonumber\\
&&S_1(r)=S(r)\mathbb{S}\;.
\end{eqnarray}
%where (\ref{df4g}) is inserted into (\ref{df1}) to construct the function $\mathbb{S}$.
When (\ref{df4g}) is inserted into (\ref{df1}), the function $\mathbb{S}$ is computed, where $M$ is an integration constant associated with the mass parameter, providing
 \begin{eqnarray}\label{df4gb}
\mathbb{S}=\frac{Q_0\gamma}{6}=const.
\end{eqnarray}
If we set the cosmological constant to zero, that is, $\Lambda=0$, then $Q_0$ is defined as \[Q_0=\frac{-9\gamma\pm\sqrt{81\gamma^2+240\gamma_1}}{20\gamma_1}.\]
In the above expression the constant  $\Lambda_{eff}$  is given by
 \begin{eqnarray}
 \label{Leff}
\Lambda_{eff}=\frac{1}{24\gamma},
\end{eqnarray}
 and it's evident that it serves as a cosmological constant.  The main finding is that we get an effective cosmological constant from the adjustment that $f(Q)$ introduces, despite the lack of the cosmological constant.  Therefore, intriguingly, the framework of $f(Q)$ gravity results in an effective cosmological constant, and the solution corresponds to an AdS spacetime when it is negative.  This characteristic, which involves the emergence of an effective cosmological constant as a consequence of the structure of $f(Q)$, was previously suggested to occur in $f(Q)$ gravitational theory. {  Thus solution (\ref{Leff}) is derived using the fact that $\Lambda=0$ and it has both of the dimensional parameters $\gamma$ and $\gamma_1$ which are involved in $Q_0$.}
 %However, in the present study, we demonstrate conclusively that it does indeed manifest, and furthermore, it holds true across various dimensions.

It should be emphasized that the solution mentioned earlier holds true solely when $\gamma$ is not equal to zero. This condition signifies its emergence as a result of the higher-order adjustment to traditional non-metricity, thus underscoring the significance of these modifications. In the scenario where $\gamma=\gamma_1=0$ and $\Lambda\neq0$, it follows that $\Lambda_{\text{eff}}$ is proportional to $\Lambda$, suggesting a reversion to conventional non-metricity in addition to a cosmological constant which implies a Schwarzschild-(A)dS solution. Finally, it is noteworthy that while $S(r)$ and $S_1(r)$ may vary by a constant, the metric's $g_{tt}$ and $g_{rr}$ components have identical event horizons and killings. Black hole, (\ref{df4g}),  exhibits a horizon at $M=\Lambda_{\text{eff}} r^3$ and a singularity at $r=0$.

{ When $Q\neq const.$ we get the solution of the system of differential equations (\ref{df2}) as:
\begin{align}
%&S={\frac {c_1 }{r}} \left[ {r}^{3}S_4{}^{2/3}-20{r}^{3}\gamma_1+9{r}^{3 }{\gamma}^{2}+3{r}^{3}\gamma\sqrt [3]{S_4}+60c_2\gamma_1\sqrt [3]{S_4} \right]\,,\nonumber\\
%
&S={\frac {1}{180c_1{\gamma_1\sqrt [3]{S_4}r}}}\left[{r}^{3} S_4{}^{2/3}-60{r}^{3}\gamma_1+36{r}^{3 }{\gamma}^{2}+6{r}^{3}\gamma\sqrt [3]{S_4}+180c_1c_2\gamma_1\sqrt [3]{S_4}\right]\,, \qquad S_1=S\mathbb{S}_1,\qquad  \mathbb{S}_1=c_1\,,
\end{align}
where $S_3=\sqrt {2700 {\gamma_1}^{2}{\Lambda}^{2}+  \left(20-540\gamma\Lambda \right) \gamma_1-9{\gamma}^{2}+216{\gamma}^{3}\Lambda}$ and $S_4=-540\gamma_1\gamma+ 5400{\gamma_1}^{2}\Lambda+216{\gamma}^{3}+60S_3\gamma_1$.
Now we proceed  our analysis, concentrating on the particular form where
\begin{equation}\label{df5}
\Lambda=\frac{1}{18\gamma} \qquad  {\text {and}} \qquad \gamma_1=\frac{3\gamma^2}{5}.
\end{equation}
{ It should be noted that, $S(r)=S_1(r)$ holds even if $\Lambda\neq0$ in the case where $\gamma\neq0$. Exploring solutions where $g_{tt}=g^{-1}_{rr}$ has got significant attention in the literature due to their ability to exhibit the suitable signature modification in the $t$ and $r$ components that is required in order to feature an event horizon \cite{Martinez:2005di,Cisterna:2014nua,Bueno:2016lrh}. Furthermore, these solutions are preferred in solar system tests.   Specifically, when the potential vector $q(r)=0$ and by using Eq.~(\ref{df5}) we get:}
\begin{eqnarray}\label{df4}
& &  S(r)= S_1=\Lambda\,r^2-\frac{M}{r}, \qquad c_1=1\Rightarrow \mathbb{S}_1=1\qquad
\end{eqnarray}
where he mass parameter $M=-c_2$  is an integration constant and $\Lambda$ is represented in Eq. (\ref{df5})\footnote{It is important to note that the disappearance of $\gamma_1$ is due to the use of Eq. (\ref{df5}).}.}
 %\begin{eqnarray}
 %\label{Leff2}
%\Lambda_{eff}=\frac{1}{12\gamma}.
%\end{eqnarray}
We derive a cosmological constant that is effective and dependent on $\gamma$, as in the previous case and when $\gamma>0$, an AdS solution is produced. Once more, the solution's horizon of Eq.~(\ref{df4}) is located at $M=\Lambda_{eff} r^3$.

\subsection{Novel solutions for charged AdS black holes}\label{S3newsol}
{ Now we are going to solve the system of differential equations (\ref{df2}) without assuming the vanishing of $q(r)$, i.e., ${{{\mathfrak T}^{{}^{{}^{^{}{\scriptstyle{em}}}}}}}^\nu_\mu\neq 0$. As is clear, such a system consists of four non-linear differential equations in three unknowns $S(r)$, $S_1(r)$ and $q(r)$.  However, we can show that in this study the matter field equation will automatically be fulfilled once Eqs. (\ref{df2}) with
non-zero $q$ is, so the system turns out to be consistent. Therefore, such a system is a closed one and has a unique solution. The  solution of such a system has the following form:
\begin{align}\label{dff}
&S=-\frac{27r^2c_4{}^5}{2744 c_3{}^3}+\frac{c_5}{r}-\frac{15c_4{}^2}{8r^2}-\frac{45c_3c_4}{32r^{10/3}}-\frac{441c_3{}^2}{176r^{14/3}}\,, \qquad S_1=S\mathbb{S}_2\,,\nonumber\\
&\mathbb{S}_2=-\frac{5488c_3{}^3r^{16/3}}{27\gamma c_4(49c_3{}^2+14c_3c_4r^{4/3}+6c_4{}^2r^{8/3})}\,, \qquad q(r)=\frac{c_4}{r}+\frac{c_3}{r^{7/3}}+\frac{49c_3{}^2}{22c_4r^{11/2}}\,.
\end{align}
Now using the following relation between the constants
\begin{align}
c_4=\frac{\varphi}{18^{1/5}}, \qquad c_3=-\left(\frac{27\gamma\varphi{}^5}{2744}\right)^{1/3}\,,
\end{align}
in Eq. (\ref{dff}) we get
\begin{eqnarray}\label{df8}
& &
 S(r)=r^2\Lambda_{eff}-\frac{M}{r}-\frac{15\varphi{}^2}{8r^{{2}}}+\frac{135(18\gamma\varphi{}^8)^{1/3}}{448\sqrt[3]{r^{10}}}
-\frac{81(18\gamma\varphi{}^{5})^{2/3}}{704\sqrt[3]{r^{14}}},\nonumber\\
&&
S_1(r)=\mathbb{S}_3(r)S(r),
\end{eqnarray}
 with
\begin{eqnarray}
&&\mathbb{S}_3(r)= \frac{1}{\left[1-\frac{(18\gamma\varphi^2)^{1/3}}{2\sqrt[3]{r^{{4}}}}+\frac
{3(18\gamma\varphi^2)^{2/3}}{8\sqrt[3]{r^{{8}}}}
 \right]^2} ,\nonumber\\
  & &
q(r)=\frac{\varphi}{r}-\frac{3(18\gamma\varphi{}^5)^{1/3}}{14\sqrt[3]{r^{{7}}}}+\frac{9(18\gamma\varphi{}^{7/2})^{2/3}}{
88\sqrt[3]{r^{{11}}}},
\label{elpot1}
\end{eqnarray}
with
\begin{align} \Lambda_{eff}=\frac{1}{18 \gamma}\,, \,\,\, \mbox{and } \,\,\, \varphi \,\,\, \mbox{is a constant of integration}\,,
\end{align}}
%{ where the dimensional quantity $\gamma$ has been absorbed in $\varphi$.}
{It is of interest to note that solution (\ref{df8}) has no form of the dimensional quantity $\gamma_1$ because we have used Eq.~(\ref{df5}).}
In solution (\ref{df8}) $M$ and $\varphi$ represent mass and electric charge, respectively. It's important to emphasize that the solution provided above is applicable only when $\varphi\neq0$. This is because when the vector potential $q(r)$ is vanishing, we revert the solution presented in Eq~(\ref{df4g}). Therefore, the presence of the non-zero charge $\varphi$ is fundamental to the characteristics of the electromagnetic sector.

Now, let's explore the characteristics of the aforementioned solution, i.e., the one given by Eq.~(\ref{df8}). To begin with, by substituting Eq.~(\ref{df8}) into  (\ref{m2}), we derive the metric as follows:
\begin{eqnarray}
\label{metric}
  &&
  \!\!\!\!\!\!\!
  ds{}^2=\Biggl[r^2\Lambda_{eff}-\frac{M}{r}-\frac{15\varphi{}^2}{8r^{{2}}}+\frac{135(18\gamma\varphi{}^8)^{1/3}}{448\sqrt[3]{r^{10}}}
-\frac{81(18\gamma\varphi{}^{5})^{2/3}}{704\sqrt[3]{r^{14}}}\Biggr]dt^2-\frac{\left[1-\frac{(18\gamma\varphi^2)^{1/3}}{2\sqrt[3]{r^{{4}}}}+\frac
{3(18\gamma\varphi^2)^{2/3}}{8\sqrt[3]{r^{{8}}}}
 \right]^2}{r^2\Lambda_{eff}-\frac{M}{r}-\frac{15\varphi{}^2}{8r^{{2}}}+\frac{135(18\gamma\varphi{}^8)^{1/3}}{448\sqrt[3]{r^{10}}}
-\frac{81(18\gamma\varphi{}^{5})^{1/3}}{704\sqrt[3]{r^{14}}}}dr^2\nonumber\\
&&
-r^2\left(d\xi^2_1+d\xi^2_2\right)\,.\nonumber\\
&&
\end{eqnarray}
As evident, in this instance, the solution appears more intricate; nevertheless, it retains its asymptotic AdS or dS nature based on the sign of $\gamma$. It's noteworthy to highlight that the $\Lambda_{eff}$ emerges as a consequence of the electromagnetic  charge, presenting an intriguing feature. However, it's crucial to acknowledge the significance of both $\gamma$ and $\gamma_1$ in shaping the structure of the solution. Consequently, this solution does not exhibit a linear form of $f(Q)$, indicating that it lacks a limit of non-metricity or an uncharged counterpart. Additionally, this particular subclass of solutions is novel and hasn't been previously documented in the literature, primarily because of the adoption of more comprehensive forms of $f(Q)$ in the current study. Consequently, the solution expressed in Eq.~(\ref{df8}) corresponds to a newly discovered charged AdS black hole within the realm of power-law $f(Q)$ gravity.

Next, we move forward to examine the singularity characteristics of the black hole solution, which involves computing curvature and non-metricity invariants.  The non-metricity scalar is derived from Eqs.~(\ref{2l}) and (\ref{Qs}), whereas the curvature scalars are computed from the metric (\ref{metric}).  Furthermore, we conclude from looking at the solution (\ref{df8}) that it is sufficient to concentrate our analysis near the function $S_1(r)$'s roots.
After computing the invariants of GR and of non-metricity we get in order:
\begin{align}
&Q^{\alpha \beta \gamma}Q_{\alpha \beta \gamma}=\frac{F_1(r)}{729r^{\frac{4}3}\gamma^2\left( 8{r}^{{\frac {8}{3}}}+3(18\gamma\varphi{}^4)^{\frac{2}3}+5(18\gamma\varphi{}^2)^{\frac{1}3}{r}^{\frac{4}3} \right) ^{4} \left( M{
r}^{\frac{11}3}-{r}^{{\frac {20}{3}}}\Lambda- 135(18\gamma\varphi{}^8)^{\frac{1}3}{r}^{\frac{4}3}+81(18\gamma^2\varphi{}^5)^{\frac{2}3}+15\varphi^2r^{\frac{8}3}\right)}\,,\nonumber\\
&P^{\alpha \beta \gamma}P_{\alpha \beta \gamma}=\frac{F_2(r)}{243r^{\frac{4}3}\gamma^2\left( 8{r}^{{\frac {8}{3}}}+3(18\gamma\varphi{}^4)^{\frac{2}3}+5(18\gamma\varphi{}^2)^{\frac{1}3}{r}^{\frac{4}3} \right) ^{4} \left( M{
r}^{\frac{11}3}-{r}^{{\frac {20}{3}}}\Lambda- 135(18\gamma\varphi{}^8)^{\frac{1}3}{r}^{\frac{4}3}+81(18\gamma^2\varphi{}^5)^{\frac{2}3}+15\varphi^2r^{\frac{8}3}\right)}\,,
\end{align}
\begin{eqnarray}
&&R=
\frac{F_4(r)}{27r^{\frac{4}3}\gamma^2\left( {r}^{{\frac {8}{3}}}+3(18\gamma\varphi{}^4)^{\frac{2}3}-5(18\gamma\varphi{}^2)^{\frac{1}3}{r}^{\frac{4}3} \right)^{3} \left( M{
r}^{\frac{11}3}-{r}^{{\frac {20}{3}}}\Lambda- 135(18\gamma\varphi{}^8)^{\frac{1}3}{r}^{\frac{4}3}+81(18\gamma^2\varphi{}^5)^{\frac{2}3}{r}^{\frac{8}3}+15\varphi^2\gamma \right)},
 \nonumber\\
&&R^{\mu \nu}R_{\mu \nu}=\frac{F_5(r)}{243r^{\frac{8}3}\gamma^2\left( {r}^{{\frac {11}{3}}}+3(18\gamma\varphi{}^4)^{\frac{2}3}-5(18\gamma\varphi{}^2)^{\frac{1}3}{r}^{\frac{4}3} \right)^{6} \left( M{
r}^{\frac{11}3}-{r}^{{\frac {20}{3}}}\Lambda- 135(18\gamma\varphi{}^8)^{\frac{1}3}{r}^{\frac{4}3}+81(18\gamma^2\varphi{}^5)^{\frac{2}3}{r}^{\frac{8}3}+15\varphi^2\gamma \right)},
\nonumber\\
&&K\equiv  R^{\mu \nu \lambda \rho}R_{\mu \nu \lambda \rho}\nonumber\\
&&=
\frac{F_6(r)}{243r^{\frac{8}3}\gamma^2\left( {r}^{{\frac {11}{3}}}+3(18\gamma\varphi{}^4)^{\frac{2}3}-5(18\gamma\varphi{}^2)^{\frac{1}3}{r}^{\frac{4}3} \right)^{6} \left( M{
r}^{\frac{11}3}-{r}^{{\frac {20}{3}}}\Lambda- 135(18\gamma\varphi{}^8)^{\frac{1}3}{r}^{\frac{4}3}+81(18\gamma^2\varphi{}^5)^{\frac{2}3}{r}^{\frac{8}3}+15\varphi^2\gamma \right)},\nonumber\\
&&
\end{eqnarray}
while computing the non-metricity  we get:
\begin{eqnarray}
Q(r)=\frac{15 \left( -32\,{r}^{{\frac {20}{3}}}\Lambda+20\,{\varphi}^{2}\gamma{r}^{8/3}+5\,
(18\gamma{\varphi}^{8})^{1/3}{r}^{4/3}+9(18\gamma^5{\varphi}^{10})^{1/3} \right)}{{r}^{{\frac {4}{3}}}\left( {r}^{{\frac {8}{3}}}+3(18\gamma\varphi{}^4)^{\frac{4}3}-5(18\gamma\varphi{}^2)^{\frac{1}3} \right)^{2}
},
\end{eqnarray}
with $F_i(r)$, $i=1\cdots 6$  representing polynomial functions of $r$.
The singularity at $r=0$ is first demonstrated by the aforementioned invariants. The behavior of these invariants near $r=0$ is provided by $(K,R_{\mu
\nu}R^{\mu \nu}) \sim \sqrt[3]{r^{-8}}$  and  $(R,Q_{\mu
\nu \rho}Q^{\mu \nu \rho},P_{\mu
\nu \rho}P^{\mu \nu \rho},Q)\sim  \sqrt[3]{r^{-4}}$, as opposed to the linear form of $f(Q)$ formulations and GR solutions of the Einstein-Maxwell theory, which have $(K ,R_{\mu \nu}R^{\mu \nu})\sim
r^{-8}$ and $(Q_{\mu
\nu \rho}Q^{\mu \nu \rho},P_{\mu
\nu \rho}P^{\mu \nu \rho},Q)\sim   r^{-4}$. This clearly demonstrates that the singularity present in our charged solution exhibits a milder behavior compared to the singularity encountered in  GR and the STEGR in the charged scenario. Lastly, it's important to observe that even though the $g_{tt}$ and $g_{rr}$ components of the metric differ within the solution, they share identical Killing and event horizons.
\begin{figure*}
\centering
\subfigure[~Behavior of $g_{tt}$]{\label{fig:1a}\includegraphics[scale=0.4]{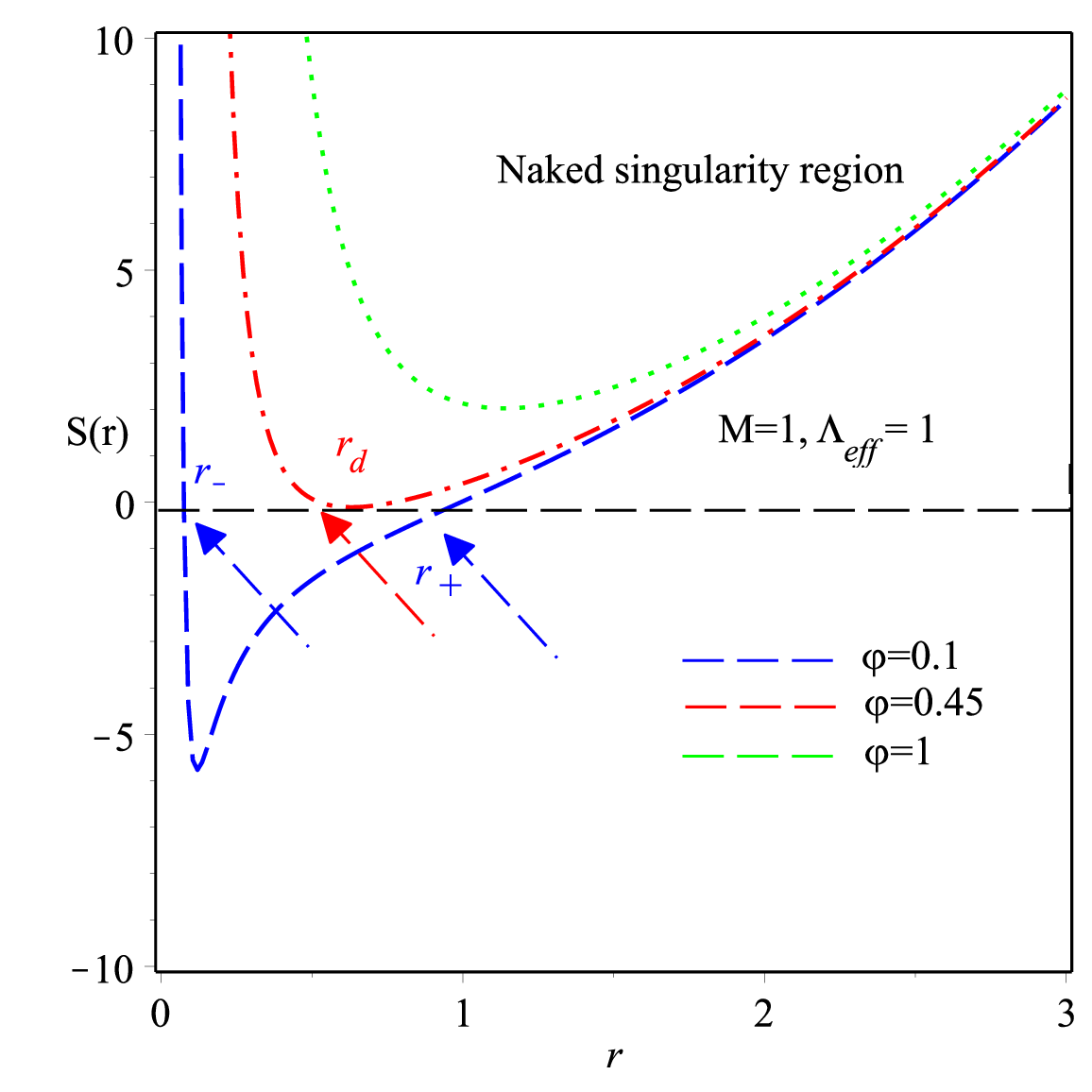}}\hspace{0.5cm}
\subfigure[~Behavior of $g_{rr}$]{\label{fig:1b}\includegraphics[scale=0.4]{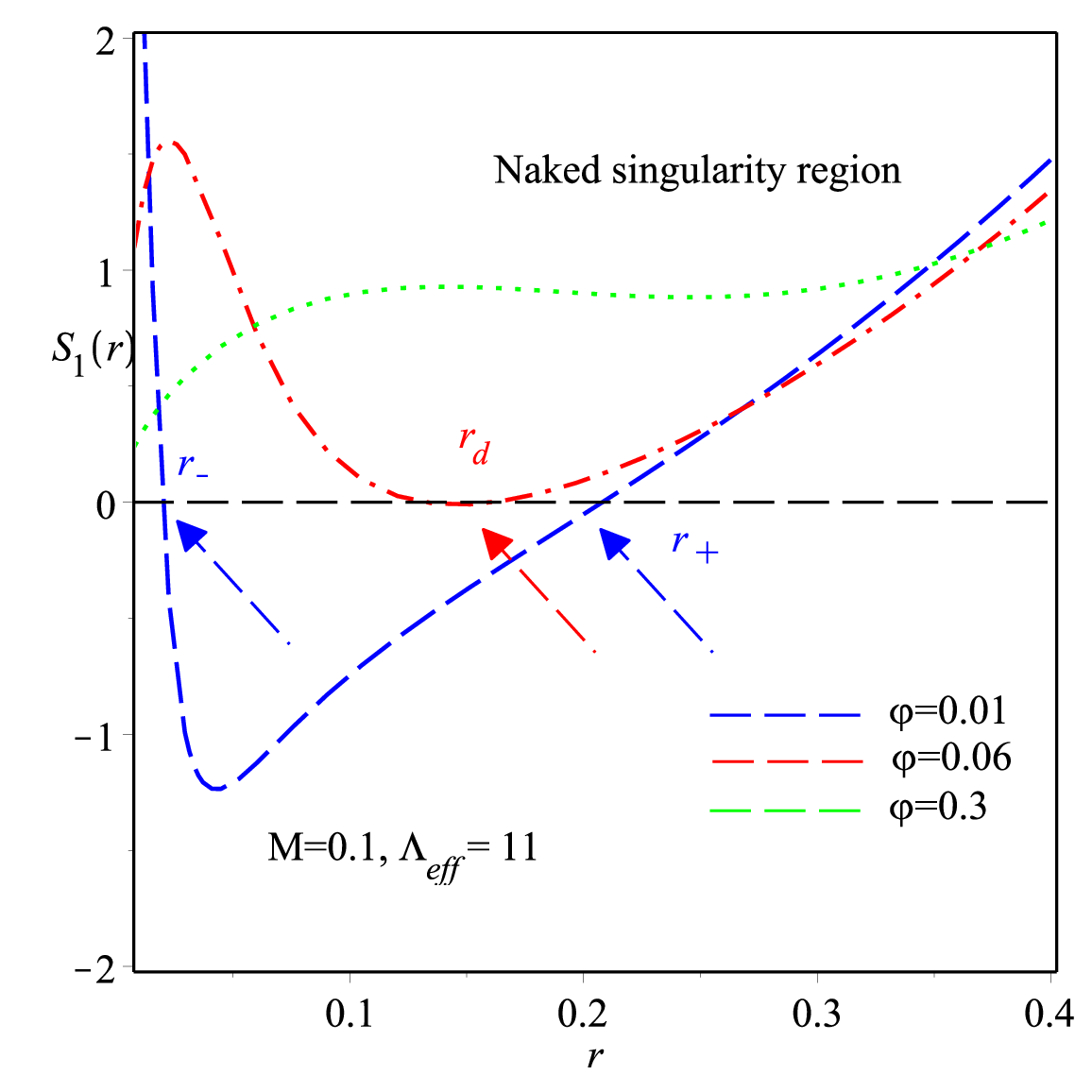}}\hspace{0.5cm}
\caption{The four-dimensional Maxwell-$f(Q)$ gravity solution's metric functions, $S(r)$ and $S_1(r)$, for different electric charge values $\varphi$, where $\kappa=1$, in the relativistic units. The outer event horizon of the black hole is represented by $r_+$, and the inner Cauchy horizon of a black hole by $r_-$. Here we put $\gamma=1$.}
\label{Fig:1}
\end{figure*}
Similar to this, we can look into the horizons of Eq. (\ref{df8}), which can also be found by looking at the solutions of $S_1(r) = 0$.
 We show in Fig. \ref{Fig:1}\subref{fig:1a}, the behavior of  $S_1(r)$ that corresponds to the solution (\ref{df8}) across different values of the model parameters.  Figure \ref{Fig:1}\subref{fig:1a} clearly illustrates the two solutions of $S_1(r)$, which match the positions of the outer event horizon $r_+$ and inner Cauchy horizon $r_-$ of the black hole \cite{Brecher:2004gn}.
It's evident that as the electric charge $\varphi$ increases and the mass parameter $M$ decreases, particularly when $\varphi=M$, we move into a parameter space without a horizon. Consequently, the naked singularity replaces the central singularity.  This represents an intriguing outcome of the charge of $f(Q)$ gravity. It's worth noting that our analysis is primarily from a mathematical perspective, and we do not delve into investigating whether such a solution could manifest physically through gravitational collapse. This particular phenomenon is absent in scenarios lacking an electromagnetic sector, as discussed previously\cite{Gonzalez:2011dr,Capozziello:2012zj}). Additionally, the two horizons combine and become degenerate for suitable values of $M$ and $\varphi$, resulting in $r_-=r_+\equiv r_{dg}$. Ultimately, to determine the horizons, we set $S_1(r)=0$, thus
\begin{equation} \label{hor11}
{M_+}=r_+\left(r_+^2\Lambda_{eff}+\frac{15\varphi{}^2}{8r_+^{{2}}}
+\frac{
45(18\gamma{\varphi}^{5})^{1/3}}{224\sqrt[3]{r_+^{10}}}
+\frac{9(18\gamma{\varphi}^{5})^{2/3}}{176\sqrt[3]{r_+^{14}}}\right).
\end{equation}

%\begin{figure}[ht]
%\centering
%\includegraphics[scale=0.4]{JFBermanosf2.eps}
%\caption{\it{{The value $m_+$ of the parameter $m$  that corresponds to the horizon
%$r_+$, of solution (\ref{df8}) of Maxwell-$f(T)$
%gravity  in four dimensions, for various values of the electric charge $q$  in units
%where $\kappa=1$.  The horizontal line at $m_+=0$  is drawn for convenience. }}
% }
%\label{Fig:2}
%\end{figure}

 \section{Maxwell-$f(Q)$ gravity with rotating black hole}\label{S5}

 We conclude this study by deriving solutions for rotating systems that satisfy the field equations within the framework of the polynomial form of $f(Q)$ gravity.  To accomplish this, we will depend on the previously extracted static solutions.
  Specifically, we will apply the subsequent two-parameter rotational transformations as:
\begin{align} \label{t1}
\bar{\xi}_{i} =-\Omega~ {\xi_{i}}+\frac{ \sigma_i}{l^2}~t,\qquad \qquad \qquad
\bar{t}=
\Omega~ t-\sum\limits_{i=1}^{2}\sigma_i~ \xi_i,
\end{align}
with $\sigma_i$ representing the rotation parameters, and where the static solution's parameter $\Lambda_{eff}$ is linked to the parameter $l$ by
\begin{eqnarray}
l=-\frac{3}{ \Lambda_{eff}}.
\end{eqnarray}
Moreover, $\Omega$ is described as
\[\Omega:=\sqrt{1-\sum\limits_{j=1}^{{2}}\frac{\sigma_j{}^2}{l^2}}.\]
%Applying the transformation (\ref{t1}) to the metric  (\ref{m2}) we obtain
 Thus, we obtain the following form for the electromagnetic potential (\ref{elpot1}):
\begin{align}
\label{Rotpot}
\bar{q}(r)=-q(r)\left[\sum\limits_{j=1}^{2} \sigma_j d\xi'_j-\Omega
dt'\right].
\end{align}
It should be noted that, despite the fact that the transformation (\ref{t1}) maintains certain local spacetime properties, it alters them globally, as demonstrated by \cite{Lemos:1994xp}, because it combines compact and noncompact coordinates.
{ The following is the representation of the metric for the transformation (\ref{t1}):
\begin{align}
\label{m1}
    ds^2=-S(r)\left[\Omega d{\bar{t}}  -\sum\limits_{i=1}^{2}  \sigma_{i}d{\bar{\xi}}
\right]^2+\frac{dr^2}{S_1(
r)}+\frac{r^2}{l^2}\sum\limits_{i=1 }^{2}\left[\sigma_{i}d{\bar{t}}-\Omega\, l^2 d{\bar{\xi}}_i\right]^2-\frac{r^2}{l^2}\left(\sigma_{1}d{\bar{\xi}}_2-\sigma_{2}d{\bar{\xi}}_1\right)^2,
\end{align}
In this case, $0\leq \xi_{i}< 2\pi$, $-\infty < t < \infty$, $0\leq r< \infty$, and $i=1,2$.} The static configuration (\ref{m2}) can be recovered as a specific instance of the generic metric stated above when the rotation parameters are set to equal zero. The coordinate transformations (\ref{t1}) can then be inverted to yield the static spacetime (\ref{metric}).

 We observe that the transformation (\ref{t1}) is locally realizable but not globally, given that the horizon's closed curves cannot be reduced to zero, the manifold's first Betti number is one \cite{Stachel:1981fg,Bonnor:1980wm}.
 %There exists a timelike Killing field $\xi=\frac{\partial}{\partial t}$ in both static (\ref{metric}) and stationary (\ref{m1}) spacetimes.  Given that the horizon's closed curves cannot be reduced to zero, the manifold's first Betti number is one \cite{Stachel:1981fg,Bonnor:1980wm}. There exists a timelike Killing field $\xi=\frac{\partial}{\partial t}$ given then by $\bar{V} =dt$ (see \cite{Bonnor_1980} for details) in both static (\ref{metric}) and stationary (\ref{m1}) spacetimes $\bar V = dt +\sigma_i d\phi_i$.  %According to De Rham's cohomology theorems, since the manifolds have a single Betti number, there are global diffeomorphisms that map the $xi$ of the two manifolds; however, there are no global diffeomorphisms that map $\bar V$ and $V$.  Because the metric turns vectors into one-forms, it means that the metrics (\ref{metric}) and (\ref{m1}) are not the same because they can only be turned into each other locally, not everywhere globally.

In power-law $f(Q)$ gravity, we have succeeded in extracting the rotating charged AdS black hole solution.  This is one of the primary findings of the current work and a novel solution. Regarding the singularity properties, the static solution of (\ref{metric}) will exhibit the same characteristics, as we can observe from the structure of (\ref{m1}).  As a result, the discussion and all of the findings in subsection \ref{S3newsol} also apply to the previously mentioned rotating solutions.  Thus, a singularity appears as $r=0$, and at the neighborhoods of $r$, the invariants are  $(K,R_{\mu
\nu}R^{\mu \nu}) \sim \sqrt[3]{r^{-8}}$  and  $(R,Q_{\mu
\nu \rho}Q^{\mu \nu \rho},P_{\mu
\nu \rho}P^{\mu \nu \rho},Q)\sim  \sqrt[3]{r^{-4}}$, unlike the charged solutions found in GR and STEGR theories.
Furthermore, the structure of the horizon exhibits qualitative similarities to the discussion presented at \ref{S3newsol}.  Interestingly, for small enough values of $\varphi$, we witness the emergence of a naked singularity.

% Interestingly, for small enough values of $\varphi$, we witness the emergence of a naked singularity. Interestingly, for small enough values of $\varphi$, we witness the emergence of a naked singularity. Interestingly, for small enough values of $\varphi$, we witness the emergence of a naked singularity. Interestingly, for small enough values of $\varphi$, we witness the emergence of a naked singularity.
%%%%%%%%%%%%%%%%%%%%%%%%%%%% Section 7 %%%%%%%%%%%%%%%%%%%%%%%%%%%%%
\section{Black hole (\ref{metric}) thermodynamics }\label{S6}
%%%%%%%%%%%%%%%%%%%%%%%%%%%%%%%%%%%%%%%%%%%%%%%%%%%%%%%%%%%%%%%%%%%%
 The definition of the Hawking temperature is  \cite{Sheykhi:2010zz,Hendi:2010gq,Sheykhi:2009pf}:
  \begin{equation}
T_+ = \frac{S'(r_+)}{4\pi},
\end{equation}
where $S'(r_+)\neq 0$ is satisfied and the event horizon is situated at $r = r_+$.
For $f({\cal Q})$ gravitational theory, the Bekenstein-Hawking entropy is given by \cite{Salako:2013gka,Bamba:2012rv}
\begin{equation}\label{ent}
\delta(r_+)=\frac{A}{4}\frac{df({\cal Q}_+)}{d{\cal Q}_+}=\frac{A}{4}\frac{df(r_+)}{dr_+}\frac{dr_+}{d{\cal Q}(r_+)},
\end{equation}
where the event horizon's area is denoted by $A$.

%The  stability of the black hole  is related to the sign of the heat capacity $C_+$.
%The   heat capacities is defined as \cite{Nouicer:2007pu,DK11,Chamblin:1999tk}
%\begin{equation}\label{m55}C_+=\frac{\partial m}{\partial r_+} \left(\frac{\partial T_+}{\partial r_+}\right)^{-1}.
%\end{equation}
%Therefore, if the heat capacity (positive, negative), then the black hole is thermodynamically (stable, unstable). Thus, a black hole that has a negative heat capacity is thermally unstable.

%To explain  the above thermodynamical quantities for the black hole solution (\ref{sol}) we begin by the constraint $b(r_+) = 0$  which  gives
%\begin{eqnarray} \label{m33}
%&& {m_+}_{{}_{{}_{{}_{{}_{\tiny Eq. (\ref{sol})}}}}}=r_+{}^{d-1}\Lambda.
%\end{eqnarray}
% Equation (\ref{m33}) indicates that the  total mass  of the black hole is a function of the horizon radius. The  relation between the function $b(r)$ and the radial coordinate $r$ is shown in  figure \ref{fig:1a} which shows the possible horizon  of the black hole solution. Also, the relation between the total mass of the black hole and the radius of the horizon is given in figure \ref{fig:1b}.

\begin{figure}[ht]
\centering
\subfigure[~Heat capacity]{\label{fig:2a}\includegraphics[scale=0.35]{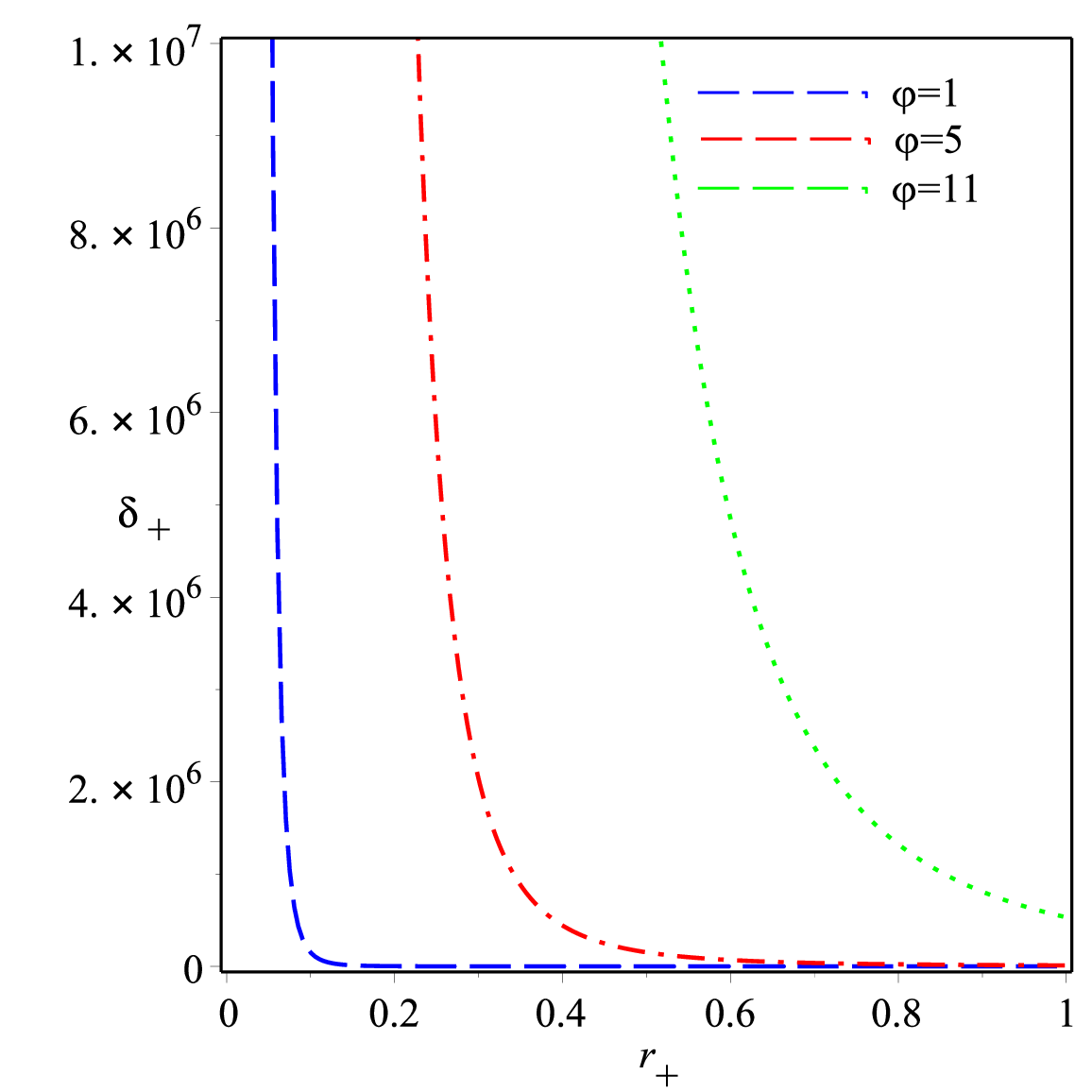}}\hspace{0.5cm}
\subfigure[~Hawking's temperature]{\label{fig:2b}\includegraphics[scale=0.35]{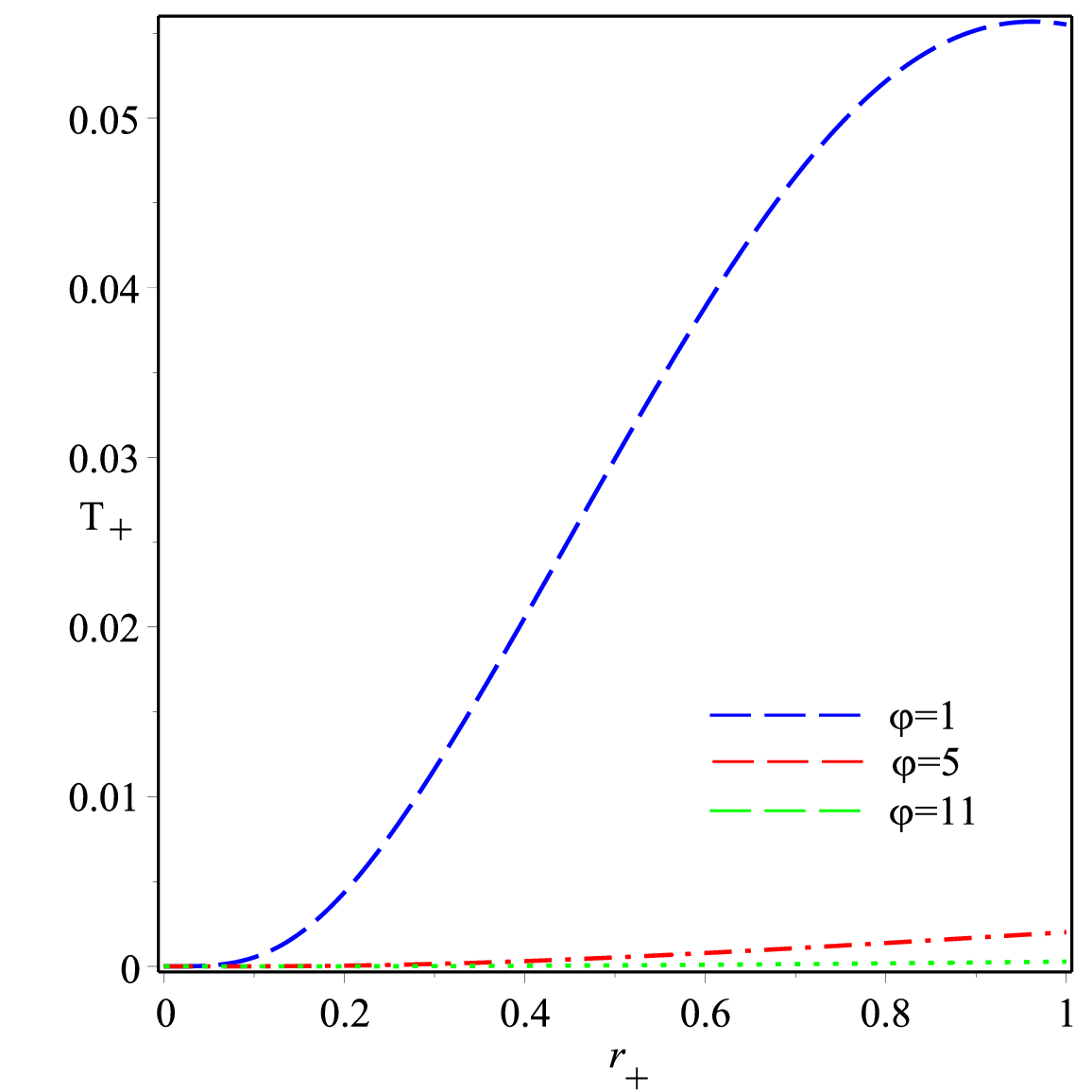}}\hspace{0.5cm}
\subfigure[~Gibb's free energy]{\label{fig:2c}\includegraphics[scale=0.35]{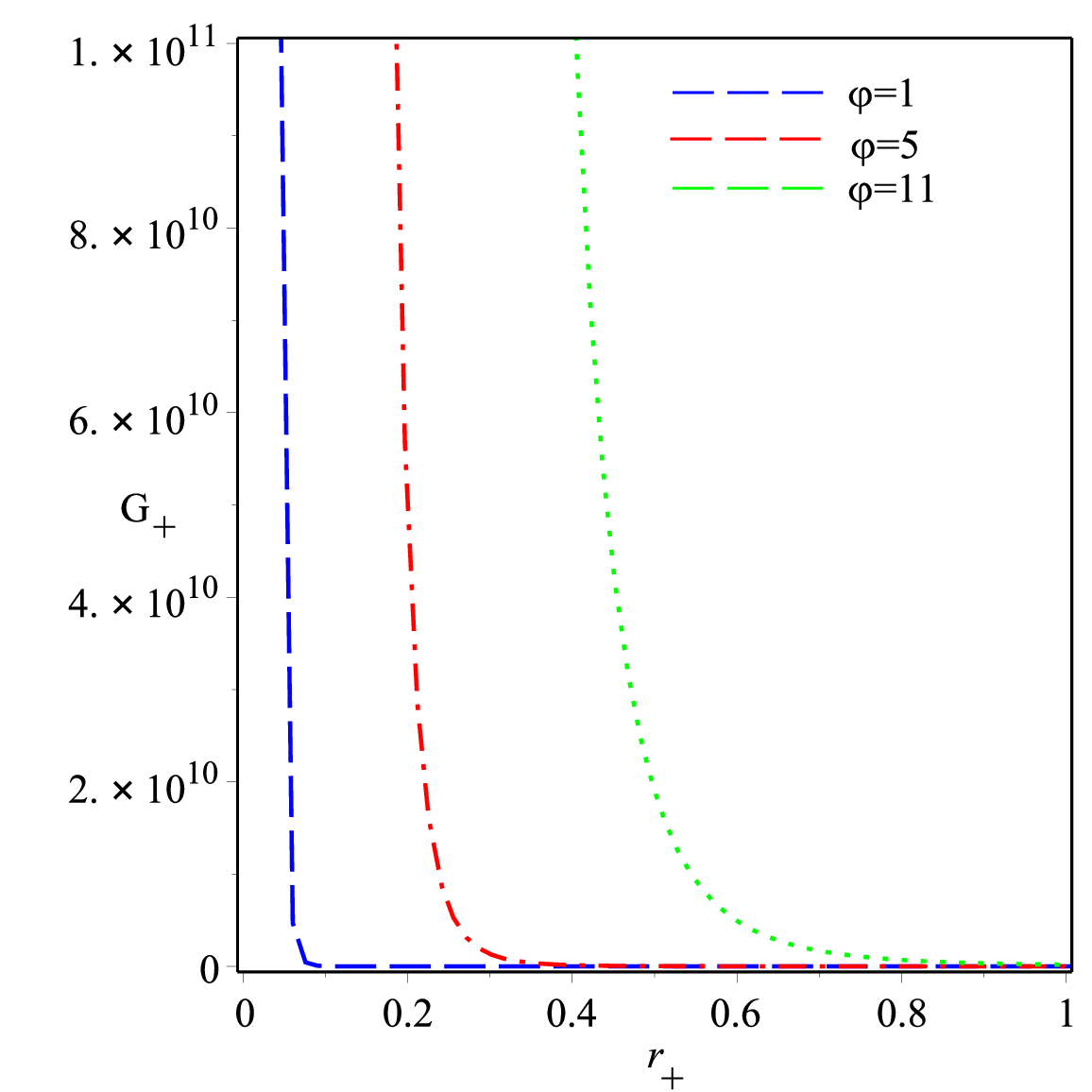}}
\caption{{\it{(a) The entropy. (b) The Hawking temperature of the black hole solution give by Eq.~(\ref{df8}). (c) The Gibb's free energy. Here we put $\gamma=1$}}}
\label{Fig:2}
\end{figure}

 %Moreover, one  can  calculate  the degenerate horizon by setting $\frac{\partial M_+}{\partial r_+}$= 0, which yields $$r_{dg} = (d-1)r_+{}^{d-2}\Lambda.$$ The relation between the degenerate horizon and the  black hole (\ref{sol}) is drawn in figure \ref{fig:2a}.
The black hole's (\ref{df8}) entropy is computed using Eq. (\ref{ent})
\begin{align} \label{ent1}
&{\delta_+}=\pi \,{r_+}^{2} \left( 1-\frac{4}3\,{\varphi}^{4} \left( 32\,{r_+}^{{\frac {20}{3}} }-135\,\sqrt [3]{18}\sqrt [3]{{\varphi}^{5}\gamma}\varphi\,\gamma\,{r_+}^{4/3} +81\,{18}^{2/3} \left( {\varphi}^{5}\gamma \right) ^{2/3}\gamma+360\,{ \varphi}^{2}\gamma\,{r_+}^{8/3} \right) {r_+}^{-4/3} \left( 3\,{18}^{2/3} \left( {\varphi}^{5}\gamma \right) ^{2/3}\right.\right.\nonumber\\
&\left.\left.-4\sqrt [3]{18}\sqrt [3]{{ \varphi}^{5}\gamma}\varphi{r_+}^{4/3}+8{\varphi}^{2}{r_+}^{8/3} \right) ^{-2}+4 /5{\varphi}^{8} \left( 32{r_+}^{{\frac {20}{3}}}-135\sqrt [3]{18} \sqrt [3]{{\varphi}^{5}\gamma}\varphi\gamma{r_+}^{4/3}+81{18}^{2/3} \left( {\varphi}^{5}\gamma \right) ^{2/3}\gamma+360{\varphi}^{2}\gamma{ r_+}^{8/3} \right) ^{2}\right.\nonumber\\
&\left. {r_+}^{-8/3}\left( 3\,{18}^{2/3} \left( {\varphi}^{5} \gamma \right) ^{2/3}-4\,\sqrt [3]{18}\sqrt [3]{{\varphi}^{5}\gamma}\varphi \,{r_+}^{4/3}+8\,{\varphi}^{2}{r_+}^{8/3} \right) ^{-4} \right)\,.
\end{align}
 Figure \ref{Fig:2}\subref{fig:2a} illustrates the entropy's behavior, displaying a positive value

The black hole solution's (\ref{df8}) yields Hawking temperature  as:
\begin{align} \label{m44}
&{T_+}=\frac{1}{4{r}^{11/3}{\pi }} \left( \frac{3}4{18}^{2/3}{\varphi}^{4/3}{\gamma}^{2/3}-\sqrt [3]{18} {\varphi}^{2/3}\sqrt [3]{\gamma}{r}^{4/3}+2{r}^{8/3} \right)  \left(\Lambda_{eff} r^2-{\frac {225}{224}}\sqrt [3]{18}{\varphi}^{8/3} {\gamma}^{1/3}{r}^{-11/3}+{\frac {189}{352}}{18}^{2/3}{\varphi}^{10/3}{ \gamma}^{2/3}\right.\nonumber\\
&\left.+{\frac {15}{4}}{\varphi}^{2}{r}^{-2}+M\right) \left( \Lambda_{eff}{r}^2-{\frac {135}{448}}\sqrt [3]{18}{\varphi}^{8/3}{\gamma}^{1/3}{r}^{ -11/3}+{\frac {81}{704}}{18}^{2/3}{\varphi}^{10/3}{\gamma}^{2/3}{r}^{ -14/3}+{\frac { 15}{8}}{\varphi}^{2}{r}^{-2}+M \right) ^{-1},
\end{align}
where the Hawking temperature at the event horizon is denoted by ${T_+}$. Figure \ref{Fig:2}\subref{fig:2b}, where we plot the Hawking temperature, displays a positive temperature.
%Now, we are going to calculate the heat capacity of the black hole solution (\ref{sol}). Using Eq. (\ref{m55}) we get
%\begin{equation} \label{enr}
%C(r_+)=4\pi r_h{}^{d-2}.
%\end{equation}
 %We plot the heat capacity in the case of 4-dimension as shown in Figure \ref{fig:3b} which shows a positive value.  Indeed, this means that the black hole solution (\ref{sol}) has  thermodynamical stability.

Gibb's free energy, which is the free energy in the grand canonical ensemble, is defined as \cite{Zheng:2018fyn,Kim:2012cma}:
\begin{equation} \label{enr}
G(r_h)=M(r_+)-T(r_+)\delta (r_+)%+P(r_+)V(r_+),
\end{equation}
%$V$ is the geometric volume  of the black hole, $P$ is the pressure which is represented by the radial    equation of (\ref{f1}), i.e. $I_r{}^r$,
where the temperature, entropy, and quasilocal energy at the event horizon are denoted by the symbols $T(r_+)$, $\delta(r_+)$, and $M(r_+)$.   We obtain Gibb's free energy by applying Eqs.~(\ref{hor11}), (\ref{ent1}), and (\ref{m44}) in (\ref{enr}) as:
\begin{align} \label{m77}
&{G_+}=-{\frac {1594323}{2560}}\, \left( {\frac {16384}{43046721}}\,\gamma\,{ r_+}^{{\frac {68}{3}}}+{\frac {10240}{129140163}}\,{r_+}^{{\frac {74}{3}}} -{\frac {71231488}{40920957}}\,{\gamma}^{3}{r_+}^{{\frac {44}{3}}}{\varphi} ^{4}-{\frac {133120}{14348907}}\,{r_+}^{{\frac {62}{3}}}{\varphi}^{2}\gamma +{\frac {16384}{4782969}}\,{r_+}^{{\frac {59}{3}}}M{\gamma}^{2}\right.\nonumber\\
&\left.-{\frac { 82632800}{10609137}}{r_+}^{{\frac {38}{3}}}{\varphi}^{6}{\gamma}^{3}+{ \frac {10935}{15488}}{r_+}^{2/3}{\varphi}^{12}{\gamma}^{6}-{\frac {48476} {567}}{\varphi}^{8}{\gamma}^{5}{r_+}^{{\frac {20}{3}}}+{\frac {5343}{154} }{\varphi}^{10}{\gamma}^{6}{r_+}^{8/3}-{\frac {77745}{4312}}{r_+}^{14/3}{ \varphi}^{10}{\gamma}^{5}+{\frac {3532015}{106722}}{r_+}^{{\frac {26}{3}} }{\varphi}^{8}{\gamma}^{4}\right.\nonumber\\
&\left.-{\frac {20480{r_+}^{{\frac {65}{3} }}M\gamma}{14348907}}-{\frac {204800}{4782969}}{\gamma}^{2}{r_+}^{{\frac {56}{3}}} {\varphi}^{2}+{\frac {189}{352}}{18}^{\frac{2}3}{\varphi}^{{\frac {34}{3}}}{ \gamma}^{{\frac {20}{3}}}+{\frac {17365760}{40920957}}{r_+}^{{\frac { 50}{3}}}{\varphi}^{4}{\gamma}^{2}+{\frac {7104640}{216513}}{\varphi}^{6}{ \gamma}^{4}{r_+}^{{\frac {32}{3}}}-\frac{4}9{18}^{\frac{2}3}{\varphi}^{{\frac {22}{3} }}{\gamma}^{{\frac {17}{3}}}M{r_+}^{5}\right.\nonumber\\
&\left.+{\varphi}^{8}{\gamma}^{6}M{r_+}^{\frac{11}3} -{\frac {2080}{243}}{r_+}^{{\frac {23}{3}}}M{\gamma}^{5}{\varphi}^{6}+{ \frac {208000}{18711}}{r_+}^{{\frac {29}{3}}}{\varphi}^{6}{\gamma}^{4}M-{ \frac {266240}{531441}}{r_+}^{{\frac {47}{3}}}M{\gamma}^{3}{\varphi}^{2}+ {\frac {323584}{59049}}{r_+}^{{\frac {35}{3}}}M{\gamma}^{4}{\varphi}^{4}\right.\nonumber\\
&\left.+ {\frac {189440}{1594323}}{r_+}^{{\frac {53}{3}}}M{\gamma}^{2}{\varphi}^{2 }-{\frac {14113280}{4546773}}{r_+}^{{\frac {41}{3}}}M{\gamma}^{3}{\varphi }^{4}-{\frac {835}{308}}{r_+}^{{\frac {17}{3}}}{\varphi}^{8}{\gamma}^{5}M +{\frac {46787380}{95482233}}{r_+}^{14}{18}^{\frac{2}3}{\varphi}^{16/3}{\gamma} ^{\frac{8}3}-{\frac {19167040}{40920957}}{r_+}^{{\frac {46}{3}}}\sqrt [3]{18 }{\varphi}^{\frac{14}3}{\gamma}^{\frac{7}3}\right.\nonumber\\
&\left.-{\frac {13059}{2464}}\,\sqrt [3]{18}{\varphi }^{{\frac {32}{3}}}{\gamma}^{{\frac {19}{3}}}{r_+}^{4/3}-{\frac {183620} {137781}}\, \left( r_+ \left( {\frac {528520}{302973}}\,r_++M \right) { \varphi}^{16/3}+{\frac {42236727}{15552614}}\,{\varphi}^{{\frac {22}{3}}} \right) {18}^{2/3}{r_+}^{10}{\gamma}^{11/3}-{\frac {20480}{4782969}}\,{ 18}^{2/3}{r_+}^{17}\right.\nonumber\\
&\left. \left(  \left( -{\frac {8}{15}}\,r_++M \right) {r_+}^{2} {\gamma}^{5/3}-1/18\,{\gamma}^{2/3}{r_+}^{5}-{\frac {24}{5}}{\gamma}^{ 8/3}M \right) {\varphi}^{4/3}+{\frac {7752320}{40920957}} \left( -{ \frac {142296}{60565}}{\gamma}^{11/3}M+ \left( -{\frac {11266}{ 109017}}\,{\gamma}^{5/3}{r_+}^{3}\right.\right.\right.\nonumber\\
&\left.\left.\left.+ \left( {\frac {38152}{181695}}\,r_++M \right) {\gamma}^{8/3} \right) {r_+}^{2} \right) {18}^{2/3}{r_+}^{13}{ \varphi}^{10/3}-{\frac {5990}{693}}\,{18}^{2/3} \left( {\frac {80919}{ 1686784}}\,{\varphi}^{{\frac {34}{3}}}+{r_+}^{2}{\varphi}^{{\frac {28}{3}}} \right) {r_+}^{2}{\gamma}^{{\frac {17}{3}}}+{\frac {2960}{2187}}\, \left( {\varphi}^{16/3}M{r_+}^{3}\right.\right.\nonumber\\
&\left.\left.+{\frac {32913}{45584}}\, \left( {\frac { 1850072}{164565}}\,r_++M \right) r_+{\varphi}^{{\frac {22}{3}}}+{\frac { 43815735}{14039872}}\,{\varphi}^{{\frac {28}{3}}} \right) {18}^{2/3}{r_+}^{ 6}{\gamma}^{14/3}-{\frac {16384}{1594323}}\, \left(  \left( 1/9\,{r_+}^{ {\frac {64}{3}}}-{\frac {5}{18}}\,{r_+}^{{\frac {61}{3}}}M \right) { \gamma}^{4/3}\right.\right.\nonumber\\
&\left.\left.+{\frac {5}{324}}\,\sqrt [3]{\gamma}{r_+}^{{\frac {70}{3}}} +{\gamma}^{7/3}{r_+}^{{\frac {55}{3}}}M \right) \sqrt [3]{18}{\varphi}^{2/3 }+{\frac {15}{44}}\, \left( {\frac {2816}{405}}\,{\varphi}^{{\frac {20}{3 }}}{r_+}^{{\frac {19}{3}}}M+ \left( {\frac {50840}{567}}\,{r_+}^{16/3}+{r_+} ^{13/3}M \right) {\varphi}^{{\frac {26}{3}}}+{\frac {352137}{34496}}\,{ \varphi}^{{\frac {32}{3}}}{r_+}^{10/3} \right) \right.\nonumber\\
&\left.\sqrt [3]{18}{\gamma}^{16/3} -{\frac {6280}{1701}}\, \left( {\frac {17248}{21195}}\,{\varphi}^{14/3}{r_+ }^{{\frac {31}{3}}}M+ \left( {\frac {119464}{21195}}\,{r_+}^{{\frac {28} {3}}}+{r_+}^{{\frac {25}{3}}}M \right) {\varphi}^{{\frac {20}{3}}}+{\frac { 683763}{193424}}\,{\varphi}^{{\frac {26}{3}}}{r_+}^{{\frac {22}{3}}} \right) \sqrt [3]{18}{\gamma}^{13/3}+{\frac {10697920}{4546773}}\, \right.\nonumber\\
&\left.\left(  \left( {\frac {168596}{167155}}\,{r_+}^{{\frac {40}{3}}}+{r_+}^{{ \frac {37}{3}}}M \right) {\varphi}^{14/3}+{\frac {26014431}{10296748}}\,{ \varphi}^{{\frac {20}{3}}}{r_+}^{{\frac {34}{3}}} \right) \sqrt [3]{18}{ \gamma}^{10/3}+{\frac {94720}{177147}}\, \left( {\gamma}^{10/3}{r_+}^{{ \frac {43}{3}}}M+{\frac {1135}{41958}}\,{\gamma}^{4/3}{r_+}^{{\frac {58} {3}}}\right.\right.\nonumber\\
&\left.\left.-{\frac {2725}{9324}}\,{\gamma}^{7/3} \left( -{\frac {304}{2725}} \,{r_+}^{{\frac {52}{3}}}+{r_+}^{{\frac {49}{3}}}M \right)  \right) \sqrt [3]{18}{\varphi}^{8/3} \right) {r_+}^{-13/3}{\gamma}^{-1} \left( -3/4\,{18} ^{2/3}{\varphi}^{4/3}{\gamma}^{2/3}+\sqrt [3]{18}{\varphi}^{2/3}\sqrt [3]{ \gamma}{r_+}^{4/3}-2\,{r_+}^{8/3} \right) ^{-4} \nonumber\\
&\left( -1/18\,{r_+}^{{\frac {20}{3}}}-{\frac {135}{448}}\,\sqrt [3]{18}{\varphi}^{8/3}{\gamma}^{4/3}{ r_+}^{4/3}+{\frac {81}{704}}\,{18}^{2/3}{\varphi}^{10/3}{\gamma}^{5/3}+{ \frac {15}{8}}\,{\varphi}^{2}\gamma\,{r_+}^{8/3}+M\gamma\,{r_+}^{11/3} \right) ^{-1}\,.
\end{align}

Figure \ref{Fig:2}\subref{fig:2c} illustrates the behaviors of the black hole's Gibb's free energy (\ref{df8}). Gibb's energy for the black hole solution (\ref{df8}) is always positive, indicating that it is more globally stable, as figure \ref{Fig:2}\subref{fig:2c} illustrates.

 \section{Conclusions}
 \label{S7}

Since the construction of the AdS/CFT procedure, there has been a growing interest in discovering and studying black hole solutions in AdS space. Apart from examining standard gravity, researchers are exploring various modifications of gravity and scenarios involving the Maxwell sector. However, most studies still fall within the realm of modified gravity that alters curvature.  In this study, we focus on a static and rotating solution with charge in 4 dimensions within the framework of $f(Q)$ gravitational theory. We emphasize the power-law assumption, which aligns with observations and is considered the most beneficial.

We demonstrate that when the charge is absent, we can derive an AdS solution \cite{DAmbrosio:2021zpm,sym16020219}. The cosmological constant in this solution is determined by the modifications introduced by $f(Q)$. We analytically derive static solutions with charge, which approach asymptotically AdS when the Maxwell sector is active. The effective cosmological constant varies based on the electric charge and the parameters of the $f(Q)$ modification. This type of solution does not have an uncharged version or a non-metricity (i.e., the linear form of $f(Q)$) or GR limit. It represents a new solution in $f(Q)$ gravity characterized by a power-law behavior, where changes in non-metricity and the inclusion of the Maxwell sector are the primary factors determining its characteristics. Moreover, we demonstrate that the potential $q(r)$ in the solution we found depends on both a higher-order electromagnetic potential and a monopole. This indicates that we could not find a charged solution with just a monopole within the framework of $f(Q)$ gravity.

Next, we investigated the behavior of the singularity of the black hole by deriving various measures of curvature and non-metricity invariants. We discovered that there is a singularity at $r=0$, but its severity is mitigated compared to standard GR due to the structure of $f(Q)$. Furthermore, we analyzed the horizons of the black holes and observed that they have a cosmological horizon as well as the evenan eventizon. However, for sufficiently large electric charges, a naked singularity emerges.

Furthermore, we derived a rotating AdS solution in Maxwell-$f(Q)$ gravity by using the analysis from the static solution and applying suitable changes. The sources of the effective cosmological constant are, as in the static case, the $f(Q)$ and the charged terms. Moreover, in the uncharged case, the singularity and horizon properties remain the same. This solution represents a new type and does not have an uncharged version or a non-metricity counterpart.

Exploring the AdS/CFT correspondence within the framework of non-metricity could greatly benefit from extracting AdS black holes in $f(Q)$ gravity. Comparing this approach to the traditional formulation based on curvature, we anticipate advantages. It would be beneficial if we could derive regular black holes within the $f(Q)$ gravitational theory. This assignment will be addressed in a separate study.

%%%%%%%%%%%%%%%%%%%%%%%%%%%%%%%%%%%%%%%%%%%%%%%%%%%%%%%%%%%%%%%%%%%%%%%%%%%%%%%%%%%%%%
%\bibliographystyle{apsrev}
%\bibliography{JRPHSRef}
%%%%%%%%%%%%%%%%%%%%%%%%%%%%%%%%%%%%%%%%%%%%%%%%%%%%%%%%%%%%%%%%%%%%%%%%%%%%%%%%%%%%%%

\end{document}